\input{aipcheck}

\documentclass[
    ,final            
  ]
  {aipproc}

\layoutstyle{6x9}

\begin{document}

\title{AdS boundary conditions and the 
Topologically Massive Gravity/CFT
correspondence}


\author{Kostas Skenderis${}^{a,b}$, Marika Taylor${}^a$ and Balt C. van Rees${}^a$}{address={${}^a$ Institute for Theoretical Physics, University of Amsterdam,\\
Valckenierstraat 65, 1018 XE Amsterdam, The Netherlands
\vskip .2cm
${}^b$ Korteweg-de Vries Institute for Mathematics,\\
Science Park 904, 1098 XH Amsterdam, The Netherlands}
}



\begin{abstract}

The AdS/CFT correspondence provides a new perspective on 
recurrent questions in General Relativity such as the 
allowed boundary conditions at infinity and the definition of gravitational 
conserved charges. Here we review the main insights 
obtained in this direction over the last decade and apply the new 
techniques to Topologically Massive Gravity. We show that this theory 
is dual to a non-unitary CFT for any value of its parameter $\mu$ 
and becomes a Logarithmic CFT at $\mu =1$.

\end{abstract}

\maketitle


\section{Introduction}

Three dimensional gravity offers an interesting arena to investigate
both the quantization of gravitational theories and holography. Since
{Einstein gravity} in {three dimensions} does not have {propagating
degrees of freedom} it is not a good toy model for higher dimensional
gravitational theories.  Adding higher derivative terms gives
propagating degrees of freedom but the theory generically then
contains {ghost-like excitations}. In recent times there has been
renewed interest in topologically massive gravity with a negative 
cosmological constant in three dimensions: 
\cite{Deser:1981wh,Deser:1982vy}
\begin{equation} 
S = \int d^3 x \left ( \sqrt{-g} (R - 2 \Lambda) + \frac{1}{2
{\mu}} { (\Gamma d \Gamma + \frac{2}{3} \Gamma^3)} \right) \label{tmg2} 
\end{equation} 
This theory admits asymptotically AdS solutions and has been used as an
arena to explore holography. It has also been conjectured to be free of instability
problems for $\mu = 1$. At $\mu \neq 1$ the perturbative massive
modes around the AdS background have negative energy and the theory is
unstable, but it was claimed in \cite{Li:2008dq} that at $\mu =1$ there are no
negative energy modes and the theory is stable. The corresponding dual
two dimensional field theory was conjectured to contain only a right moving sector, and
thus to be a chiral conformal field theory. 

This claim proved controversial as other authors found non-chiral
modes and instabilities at $\mu =1$
\cite{AyonBeato:2004fq,AyonBeato:2005qq,Carlip:2008jk,Grumiller:2008qz,Park:2008yy,Grumiller:2008pr,Carlip:2008eq,Carlip:2008qh,Giribet:2008bw,Blagojevic:2008bn}.
The unstable modes have fall-off conditions which are different from those 
that the metric satisfies in pure three-dimensional Einstein gravity, 
the so-called Brown-Henneaux boundary conditions \cite{Brown:1986nw} for
asymptotically AdS spacetimes. The main issue is then 
the question: {\it What are the allowed fall-off
  conditions for the fields at infinity?}

The traditional point of view regarding fall-off conditions goes back (at least)
to the work of Regge and Teitelboim \cite{Regge:1974zd} and can be 
summarized as follows:
\begin{enumerate}
\item Select {physically "reasonable"} fall-off conditions such that relevant solutions, for example
black holes, satisfy them.
\item Check that conserved charges are {finite} with this choice.
\end{enumerate}
One may then consider different fall-off conditions as defining {different theories}.

The AdS/CFT correspondence however provides a {\bf new perspective} which leads to a
{comprehensive answer} to such questions. The focus of this article will be to explain the new insights and {methodology}
originating from AdS/CFT and their applications to topologically massive gravity. More details can be found in
the article \cite{Skenderis:2009nt}.

\section{AdS/CFT: basics}

An asymptotically $AdS$ spacetime has a conformal boundary at which
boundary conditions for all bulk fields need to be defined. In the
framework of the AdS/CFT correspondence the bulk fields $\phi^I_{(0)}$
parametrizing these boundary conditions at conformal infinity are
identified with sources that couple to operators $O^I$ of the dual
CFT. The defining relation of the AdS/CFT correspondence is that the 
on-shell action, $S_{onshell}[ \phi_{(0)}]$, is the generating functional of CFT correlation
functions:
\begin{equation}
\langle {O} \rangle \sim \frac{\delta S_{onshell}[{\phi_{(0)}}]}{\delta {\phi_{(0)}}}, \quad
\langle {O(x) O(y)} \rangle \sim 
\frac{\delta^2 S_{onshell}[{\phi_{(0)}}]}{\delta {\phi_{(0)}}(x)\delta {\phi_{(0)}}(y)},
\end{equation}
etc.
These identifications lead to new intuition 
about the boundary conditions.
In quantum field theory the sources that couple to operators are unconstrained, because one
functionally differentiates with respect to them.
This implies that one should be able to formulate the bulk/boundary problem by specifying
arbitrary functions/tensors as boundary conditions for bulk fields.

Let us consider the case where the bulk field of interest is the {metric}.
In the physics literature, prior to the AdS/CFT correspondence,
there were a number of works
discussing { Asymptotically AdS spacetimes}, for example 
\cite{Abbott:1981ff,Ashtekar:1984zz,Henneaux:1985tv}.
In these works the metric approaches that of AdS at conformal infinity; 
the spacetimes are asymptotically AdS. 

For AdS/CFT however such boundary conditions do not suffice: one needs 
{more general boundary conditions}. In particular, the boundary conditions must
be parametrized by an {unconstrained metric}, since this metric should act as a source for the
{ energy momentum tensor $T_{ij}$} of the dual CFT.
Fortunately this more general set-up has been developed in the
mathematics literature \cite{FeffermanGraham}. The corresponding spacetimes
are called {Asymptotically locally AdS spacetimes} (AlAdS).

An AlAdS spacetime always admits the following
metric in a finite neighborhood of the conformal boundary, located at {$r=0$}:
\begin{equation}
ds^2 = \frac{dr^2}{r^2} + \frac{1}{r^2} g_{ij}(x,r) dx^i dx^j
\end{equation}
where
\begin{equation}
\lim_{r \to 0} g_{ij}(x,r) = {g_{(0)ij}(x)}
\end{equation}
is an {arbitrary non-degenerate} metric. The coordinates used here are Gaussian normal
coordinates centered at the conformal boundary.

Let us emphasize that the only requirement put on $g_{ij}(x,r)$ {a priori}
is that it should have a {non-degenerate} limit as {$r \to 0$}.
The precise form of $g_{ij}(x,r)$ is determined by {solving} the bulk
field equations {asymptotically}.
This problem reduces to solving {algebraic} equations, so the {{\bf most general}
asymptotic solution} can be readily found for any given bulk theory that admits AlAdS solutions.

For {Einstein gravity} in $(d+1)$ dimensions, the relevant expansion is \cite{FeffermanGraham,Henningson:1998gx}:
\begin{equation}
g_{ij}(x,r) = g_{(0)ij}(x) + r^2 {g_{(2)ij}} + \cdots
+ r^d (g_{(d)ij} +  {h_{(d)ij}} \log (r^2)) + \cdots
\end{equation}
Here the {coefficients} $(g_{(2) ij}, \cdots, h_{(d)ij})$ are locally determined
in terms of $g_{(0)}$. $g_{(d)ij}$ is only partially determined by asymptotics: this coefficient is related
via AdS/CFT to the {1-point function of $T_{ij}$} and thus to {bulk conserved charges}. The trace
and divergence of $g_{(d)ij}$ are determined and relate to dilatation and diffeomorphism Ward identities.
The logarithmic coefficient $h_{(d)}$ is non-zero when $d$ is even and is greater than two, 
and it is related to the Weyl anomaly of the boundary
theory \cite{deHaro:2000xn}
\begin{equation}
{h_{(d)} \sim  \frac{\delta}{\delta g_{(0)}} \int (conformal\ anomaly)}.
\end{equation}
Note that in the specific case of pure {Einstein gravity} in three bulk dimensions
\begin{equation} \label{d2FG}
g_{ij}(x,r) = g_{(0)ij}(x) + r^2 g_{(2)ij} + \cdots
\end{equation}
In this case, $h_{(2)}$ actually vanishes because the integral of the conformal anomaly is a
topological quantity (the Euler number).

The precise form of this expansion is {specific to Einstein gravity}.
Coupling to matter changes the coefficients.
For example, coupling Einstein gravity to a free massless scalar 
induces a {logarithmic term} in the expansion, i.e. $h_{(2)} \neq 0$ in this theory \cite{deHaro:2000xn}.
There is also 
an example of 3d gravity coupled to scalars with $\log^2$ terms in the asymptotic expansion, see 
\cite{Kanitscheider:2006zf}, appendix E. Even the power of the leading order 
correction can change, for example it can be $r$ rather 
than $r^2$ \cite{Berg:2001ty}. 

Note that the {Brown-Henneaux} boundary conditions are as in (\ref{d2FG}) with
the additional restriction {$g_{(0)ij}(x) = \delta_{ij}$ (in the Euclidean), i.e.
the metric is asymptotically $AdS_3$. 
Moreover, these boundary conditions are often
quoted as:
\begin{equation}
g_{ij}(x,r) = \delta_{ij} + {\cal O}(r^2),
\end{equation}
i.e. it is assumed that the fall-off of the subleading terms is polynomial rather than logarithmic. 
It is important to emphasize that
in the AdS/CFT correspondence such boundary conditions are not sufficiently general. For example, 
the Brown-Henneaux boundary conditions are violated whenever one wishes to consider the CFT in a non-trivial background, or
when one wishes to compute correlation functions of the stress energy tensor. Logarithmic terms generically
arise in the expansion of the subleading terms and are related to matter and gravitational conformal anomalies.

\subsection{Conserved charges}

This is an another area where the AdS/CFT duality provides a new and systematic
approach \cite{Skenderis:2000in}. In quantum field theory the energy is computed using the {energy momentum tensor},
\begin{equation}
E=\langle H \rangle = \int d^{d{-}1}\!x \langle T_{00} \rangle
\end{equation}
Generically this expression needs {renormalization} due to {UV infinities}.

In the AdS/CFT correspondence
\begin{equation}
\langle T_{ij} \rangle = \frac{\delta S_{onshell}[{ g_{(0)}}]}{\delta { g_{(0)}^{ij}}}
\end{equation}
This expression is also formally { infinite}, 
due to the infinite volume of spacetime
({IR divergences}) and needs {\it holographic renormalization} 
\cite{Skenderis:2002wp}.

One can holographically renormalize the theory by adding 
{local boundary covariant counterterms}
\cite{Henningson:1998gx,Henningson:1998ey,Balasubramanian:1999re} and thus 
obtain a {finite 1-point function for $T_{ij}$} for a general AlAdS
spacetime \cite{deHaro:2000xn}
\begin{equation}
\langle T_{ij} \rangle \sim g_{(d)ij} + X_{ij}[g_{(0)}]
\end{equation}
with $X_{ij}[g_{(0)}]$ a known local function of $g_{(0)}$.
One can furthermore prove {rigorously} from first principles (e.g. using Noether's method or Wald's covariant
phase space methods) that the {holographic charges} are the correct {gravitational
conserved charges} \cite{Papadimitriou:2005ii}. Note that the proofs given in \cite{Papadimitriou:2005ii} apply equally well to
cases where there are logarithmic terms in the asymptotic expansions. 

\subsubsection{Summary}

In summary, the {holographic  methodology} that replaces previous approaches is:

\begin{enumerate}

\item {Derive the most general solution} of the bulk equations
with {general Dirichlet boundary conditions} for all fields.

\item General results guarantee that the {conserved charges} are well-defined
and can be obtained from the {holographic 1-point functions}.

\end{enumerate}

The holographic framework allows one to go further and obtain new information by computing
{two and higher point functions}.

\section{Application to TMG}

Topologically massive gravity is obtained by adding to 3d Einstein gravity the
gravitational Chern-Simons term, see equation (\ref{tmg2}). The equations of motion are:
\begin{equation}
R_{\kappa \lambda} + 2 g_{\kappa \lambda} + \frac{1}{\mu} 
\epsilon_{\kappa}^{\phantom{\kappa} \rho \sigma} \nabla_{\rho} R_{\sigma \lambda} + 
\kappa \leftrightarrow \lambda= 0, \nonumber
\end{equation}
and these admit {asymptotically  $AdS$} solutions, for example the BTZ black hole, as well
as {perturbative massive modes} when one expands around $AdS$. When $\mu \neq 1$ however, 
the massive modes have negative energy and the theory is known to be unstable.

In exploring holography for TMG 
Strominger et al \cite{Li:2008dq} claimed that the dual 2d CFT is chiral at $\mu =1$ in the following sense:
\begin{enumerate}
\item{There are {no left moving modes in the bulk} satisfying Brown-Henneaux boundary conditions.}
\item{The {left moving central charge} $c_{L}$ of the CFT is zero at $\mu = 1$.}
\item{There are no negative energy modes and the theory is therefore {stable} at $\mu = 1$.}
\end{enumerate}
A holographic correspondence between TMG and a {chiral} CFT was proposed.

However, the non-chiral mode of topologically massive gravity found in \cite{Grumiller:2008qz}
has the asymptotic form
\begin{equation}
g_{ij}(x,r) = \delta_{ij} + r^2 (g_{(2)ij} + \log (r^2) h_{(2)ij}) + \cdots
\end{equation}
which differs from the Brown-Henneaux boundary conditions because of the $h_{(2)}$
logarithmic term. A discussion followed as to whether such boundary conditions could be
{consistent} and subsequently it was proven by \cite{Henneaux:2009pw}
that conserved charges are indeed finite with such boundary conditions. 
As mentioned above, the fact that the charges are finite is unsurprising 
since the general proof given in \cite{Papadimitriou:2005ii}, 
although strictly speaking not applicable for TMG,
encompasses cases with logarithmic fall-off behaviors. 

From the perspective of AdS/CFT:

\begin{enumerate}

\item a subleading log is not surprising, as the subleading coefficients {\it routinely change and involve logs} as
one changes the bulk action;

\item the form of the asymptotic expansion should not be {fixed by hand}
but should rather be {derived} by solving the bulk equations asymptotically.

\end{enumerate}

We will return to the most general asymptotic solution of TMG shortly.

\section{Holography for TMG and LCFT}

We now move to apply holographic methodology to the topologically massive gravity.
Let us first consider the theory at $\mu =1$.
There is an important new element compared to earlier holographic literature:
the field equations are {third order} in derivatives, so there
are two independent boundary data: one can fix the {metric} and a certain 
component of 
the {extrinsic curvature}. 
The boundary metric {$g_{(0)ij}$} is the source for the {energy momentum tensor $T_{ij}$}.
The boundary field {$b_{(0)ij}$} parametrizing the boundary behavior of  
the extrinsic curvature is a {source for
a new operator {$t_{ij}$}}.

We need one further ingredient. It turns out that $t_{ij}$ is obtained
as a limit of an {irrelevant} operator.
{In CFT}, when one couples an irrelevant operator, this generates
{severe UV divergences} and the theory is {not conformal in the UV}.
{In gravity}, a source for an irrelevant operator introduces
{severe IR divergences} and the solution is {not asymptotically AdS}
\cite{deHaro:2000xn}.
In both cases, one bypasses the problems by treating the source
{perturbatively} and thus 
we will work to {first order in $b_{(0)}$}, which suffices for the computation
of correlation functions that involve at most two insertions of $t_{ij}$.
In particular, we can compute all 2-point functions.

The {most general asymptotically locally AdS solution} (i.e. with non-degenerate conformal boundary) 
of the TMG equations of motion, with terms linear in the source $b_{(0)}$ for the irrelevant operator 
also included, is then: 
\begin{equation}
ds^2 = \frac{dr^2}{r^2} + \frac{1}{r^2} g_{ij}(x,r) dx^i dx^j
\end{equation}
with
\begin{equation}
g_{ij}(x,r) = {b_{(0)ij}} \log r^2  +  g_{(0)ij}
+ r^2 ( g_{(2)ij} + {b_{(2)ij}} \log r^2 ) + \cdots
\end{equation}
Only {$b_{(0)\bar{z} \bar{z}}$} is non-zero and is the source for the {new operator $t_{zz}$}.
The subleading coefficients $g_{(2)}$ and $b_{(2)}$ are constrained partially by the
asymptotic analysis, with the constraints as usual relating to Ward identities. 

The (finite) holographic 1-point functions can be computed in {complete generality}:
\begin{eqnarray}
\langle T_{ij} \rangle &=&
\frac{1}{4 G_N}\Big(g_{(2)ij} + \frac{1}{2} R[g_{(0)}] g_{(0)ij} \\
&&- \frac{1}{2}\Big(  \epsilon_i^{\phantom{i}k} g_{(2)kj} + (i \leftrightarrow j)  \Big) 
- 2 b_{(2)ij} + \frac{1}{2} A_{ij}[g_{(0)ij}] \Big) \nonumber \\
\langle t_{zz} \rangle &=& \frac{1}{2 G_N} (g_{(2)zz} + b_{(2)zz}) \nonumber
\end{eqnarray}
$T_{ij}$ satisfies the {expected anomalous CFT Ward identities}:
\begin{eqnarray}
\langle T_i^i \rangle &=& \frac{1}{4 G_N}\Big( \frac{1}{2} R[g_{(0)}] + \frac{1}{2} A_i^i[g_{(0)}] \Big)
\\
\nabla^j \langle T_{ij} \rangle &=& \frac{1}{4 G_N}\Big(\frac{1}{4} \epsilon_{ij} \nabla^j R[g_{(0)}]
+ \frac{1}{2} \nabla^j A_{ij}[g_{(0)}]\Big) \nonumber
\end{eqnarray}
The right hand side of the second equation contains the expected
consistent (non-covariant) diffeomorphism anomaly. The improved 
energy momentum tensor, $\hat{T}_{ij} = T_{ij} - \frac{1}{8 G_N} A_{ij}$ 
has instead a covariant diffeomorphism anomaly \cite{Bardeen:1984pm}. From 
the trace Ward identity one can extract the sum of left and right 
central changes, $c_L+c_R$. 

The energy momentum tensor $T_{ij}$ can be used to obtain the {conserved  charges}.
For example one can compute the conserved charges for the BTZ black hole:
\begin{eqnarray}
ds^2 &=& \frac{dr^2}{r^2} - \left [ \frac{1}{r^2} - \frac{1}{2} (r_+^2 + r_-^2) 
+ \frac{1}{4} (r_+^2 - r_-^2)^2 r^2 \right ] dt^2 \\
&& \qquad + \left [ \frac{1}{r^2} + \frac{1}{2} (r_+^2 + r_-^2) + \frac{1}{4} (r_+^2 - r_-^2)^2 r^2 \right ] d\phi^2 + 2 r_+ r_- dt d\phi. \nonumber
\end{eqnarray}
The stress energy tensor becomes {chiral} at $\mu=1$,
\begin{equation}
T_{\bar{z} \bar{z}} = \frac{2}{G_N} (r_+ + r_-)^2, \quad T_{zz}=0
\end{equation}
and the {conserved charges} are
\begin{eqnarray}
M &=& - \int d\phi T^t_t = \frac{\pi}{4 G_N} (r_+ + r_-)^2  \nonumber\\
J &=& - \int d\phi T^t_\phi = M 
\end{eqnarray}
Note that $J = M$ even away from extremality, i.e. for $r_+ \neq |r_-|$.

Given the general expressions for the 1-point functions, we can use the general solution of the {linearized equations of motion}
about AdS to extract the following non-zero 2-point functions:
\begin{eqnarray}
\langle t_{zz}(z,\bar z)t_{zz}(0) \rangle 
&=& \frac{{ (3/G_N)}  \log |z|^2}{z^4}, \\
\langle t_{zz}(z,\bar z)T_{zz}(0) \rangle  &=& \frac{{ (-3/G_N)}}{2 z^4}, \nonumber \\
\langle T_{\bar z \bar z}(z,\bar z)T_{\bar z \bar z}(0) \rangle  &=& \frac{{ (3/G_N)}}{2 \bar z^4},
\nonumber
\end{eqnarray}
These are precisely the non-zero 2-point functions of a {Logarithmic CFT} with central charges:
\begin{equation}
c_L=0, \qquad {c_R = \frac{3}{G_N}}.
\end{equation}
In the left moving sector the operators $(t_{zz},T_{zz})$ form a logarithmic pair with non-diagonalizable two point functions
and "new anomaly" parameter 
\begin{equation}
b = - \frac{3}{G_N}
\end{equation}
such that 
\begin{equation}
\langle t_{zz}(z,\bar z) T_{zz}(0) \rangle  = \frac{b}{2 z^4}, 
\end{equation}
characterizing the LCFT. 

We also analyzed the theory in the neighborhood of $\mu=1$. 
Letting $\mu = 2 \lambda + 1$, near $\lambda = 0$, the general solution to the
linearized equations of motion is expanded near the boundary as
\begin{equation}
h_{ij} = h_{(-2\lambda)ij} r^{-2 \lambda} + h_{(0)ij} + h_{(2)ij} r^2  + \ldots,
\end{equation}
where $h_{(0) ij}$ is the usual source for the energy-momentum tensor
and $h_{(-2 \lambda) ij}$ is traceless and chiral and acts as a source for an {irrelevant operator $X_{ij}$}.

The nonvanishing {two-point functions} are:
\begin{eqnarray}
\langle T_{\bar z \bar z}(z,\bar z) T_{\bar z \bar z}(0) \rangle &=&
\frac{3}{ 2 G_N}\frac{\lambda + 1}{2\lambda + 1} \frac{1}{\bar z^4} \nonumber \\
\langle T_{zz}(z,\bar z)T_{zz}(0) \rangle &=&
\frac{3}{ 2 G_N}\frac{\lambda}{2\lambda + 1} \frac{1}{z^4} \nonumber \\
\langle X_{zz}(z,\bar z) X_{zz}(0) \rangle &=&
- \frac{1}{2 G_N} \frac{\lambda (\lambda + 1)(2\lambda +3)}{2\lambda +1}\frac{1}{z^{2\lambda + 4}\bar z^{2\lambda} }. \nonumber
\end{eqnarray}
From these expressions we see that
\begin{equation}
(c_L, c_R) =  \frac{3}{2G_N}  \Big( 1 - \frac{1}{\mu},1+ \frac{1}{\mu} \Big)
\end{equation}
whilst $X$ has weights $(h_L,h_R) = (2 + \lambda, \lambda)$.

These correlation functions {smoothly} reduce to those at {$\mu \to 1$}; the operator
$t_{zz}$ is given by
\begin{equation}
t_{zz} = - \frac{1}{\lambda} (X_{zz} - T_{zz}) 
\end{equation}
and we recover the value of $b$ given previously.
In fact, our discussion mirrors the {degeneration} of a CFT to a
logarithmic CFT as $c \rightarrow 0$ discussed by \cite{Kogan:2002mg}.
As here their logarithmic partner of the stress energy tensor originates from another primary whose dimension
approaches $(2,0)$ in the $c \rightarrow 0$ limit.
There are other ways to take a $c \rightarrow 0$ limit (avoiding "catastrophe", and demanding that the OPE
remains well defined), but it is this approach which is 
realized holographically.

From the form of the 2-point functions one finds that the CFT contains a 
{state $|X \rangle$ of negative norm} and { $\langle X| H |X \rangle <0$} in 
that state. This is the counterpart of the {bulk instability} due to {negative energy} of massive gravitons.

\section{Conclusions}

Topologically massive gravity at $\mu =1$ is dual to 
a {logarithmic CFT} and therefore it is {not unitary}.
Away from the "chiral point" the theory contains {states of negative norm}.
One may try to {restrict to the right-moving sector} of the
theory, which could yield a {consistent chiral subsector}. 
Arguments for such a truncation at the classical level were given in \cite{Maloney:2009ck}. 
From the current perspective a necessary requirement
for such a truncation would be that the logarithmic operator $t$ is not generated
in the OPE of the right-moving operators; in particular, the three point 
function {$\langle t \bar{T} \bar{T}\rangle $} must vanish. This indeed 
holds for certain LCFTs, in particular for those discussed
in  \cite{Kogan:2002mg}, and it would be interesting to compute this 
3-point function holographically for TMG at $\mu=1$.

We should emphasize however that the existence of such a truncation 
only shows that a set of operators of the LCFT (in this case
the right moving stress energy tensor) form a closed subsector, 
not that this subsector has a dual of its own. To give an example in a 
more familiar setting let us consider $N=4$ SYM in four dimensions
and the dual string theory on $AdS_5 \times S^5$. There are a number 
of consistent truncations of the bulk theory. For example, it is generally believed
(and it has been proven for certain subsectors) that the maximally supersymmetric 
$SO(6)$ gauged supergravity in five dimensions is a consistent truncation 
of type IIB supergravity on $S^5$. The
existence of this consistent truncation however does not imply that there is 
a new duality: the dual theory is always $N=4$ SYM and the consistent 
truncation only implies that certain 
operators (those in the stress energy supermultiplet in this example)
are closed under OPE's in the large N limit. 

Finally, the following argument suggests that
difficulties are generic in formulating a duality between a unitary CFT
and a bulk theory that only involves
three dimensional gravity, such as TMG, instead of a string theory 
that at low energies reduces to the gravitational theory.
In the AdS/CFT correspondence we expect to have a bulk field 
for every boundary gauge invariant operator. The existence
of black holes in these theories implies that the dual theory has a very large 
number of operators to account for the entropy of the black hole. 
For each of those operators the bulk theory should have a corresponding
bulk field. Pure gravity however only contains the metric so it can 
only describe the stress energy tensor holographically. If we 
allow for higher derivative terms, as in the case of TMG, 
which are treated exactly (rather 
than perturbatively, as they would be in a string theory set up) 
one can incorporate a few more gauge invariant operators 
but then the theory generically becomes non-unitary.
In all cases the bulk description is 
missing the fields that would provide the sources for the operators
dual to the black hole microstates. Instead, in a string theory 
set up these operators would be dual to corresponding string states.


\begin{theacknowledgments}

This work is part of the research program of the 'Stichting voor
Fundamenteel Onderzoek der Materie (FOM)', which is financially
supported by the 'Nederlandse Organisatie voor Wetenschappelijk
Onderzoek (NWO)'. The authors acknowledge support from NWO, KS via a
Vici grant, MMT via the Vidi grant "Holography, duality and time
dependence in string theory" and BvR via an NWO Spinoza grant.
\end{theacknowledgments}



\bibliographystyle{aipproc}   

\bibliography{biblio}

\IfFileExists{\jobname.bbl}{}
 {\typeout{}
  \typeout{******************************************}
  \typeout{** Please run "bibtex \jobname" to optain}
  \typeout{** the bibliography and then re-run LaTeX}
  \typeout{** twice to fix the references!}
  \typeout{******************************************}
  \typeout{}
 }

\end{document}